\documentclass[iopart,twocolumn,superscriptaddress]{revtex4-2}
\usepackage{amssymb}
\usepackage{amsmath}
\usepackage{mathrsfs}
\usepackage{graphicx}
\usepackage{float}
\usepackage[normalem]{ulem}
\usepackage{color}
\usepackage{bm}

\begin{document}
\title{Photonic corner skin modes}
\author{Weiwei Zhu}
\affiliation{Department of Physics, National University of Singapore, 117551, Singapore}
\author{Jiangbin Gong}
\email{phygj@nus.edu.sg}
\affiliation{Department of Physics, National University of Singapore, 117551, Singapore}
\affiliation{Centre for Quantum Technologies, National University of Singapore, 117543, Singapore}
\affiliation{Joint School of National University of Singapore and Tianjin University, International Campus of Tianjin University, Binhai New City, Fuzhou 350207, China}

\pagebreak

\begin{abstract}
Useful in the enhancement of light-matter interaction, localization of light is at the heart of photonics studies.   Different approaches have been proposed to localize light, including those based on dynamical localization, topological trivial or nontrivial defects in the band gap of photonic crystals, and bound states in the continuum.  Recent studies on non-Hermitian skin effect have provided us new means to localize waves. In this work, we propose a new method towards localized light, called photonic corner skin modes arising  from second-order non-Hermitian skin effect and gain-loss symmetry on a lattice.    Specifically, we propose to make use of small pseudo-inversion symmetric gain/loss, which does not close the band gap, to realize a photonic Chern insulator with chiral edge states. The chiral edge states then accumulate at certain corners of the system.  Intriguing phenomena such as  corner skin modes arising from an underlying bipolar second-order non-Hermitian skin effect and multiple-corner skin modes are predicted in continuous systems.  
\bigskip

\end{abstract}

\maketitle

\section{Introduction}
Due to the non-Hermitian skin effect (NHSE),  the spectrum of a non-Hermitian system is sensitive to its boundary condition. Under open boundary conditions (OBC), NHSE in one-dimensional systems causes all the bulk states to localize at one edge of the system~\cite{Yao2018,Lee2019a,Okuma2020,Zhang2020}.
While the NHSE itself has topological origin associated with point-gap topology~\cite{Okuma2020,Zhang2020}, it also challenged our understanding of the usual bulk-boundary correspondence of topological band theory and even led to the concept of generalized bulk-boundary correspondence,  via which the topological invariants are defined in the so-called generalized Brillouin zone~\cite{Yokomizo2019,Yang2020,Deng2019,Xiao2020,Helbig2020}.

To date the first-order NHSE has been widely studied both in theory and experiment~\cite{Yao2018,Lee2019a,Okuma2020,Zhang2020,Okuma2020,Zhang2020,Yokomizo2019,Yang2020,Deng2019,Xiao2020,Helbig2020,Hofmann2020}. However, there are much less work on the second-order NHSE, where certain states (the number of states is proportional to the length of the system) are localized at corner, with the bulk states still extended~\cite{Lee2019,Li2020,Kawabata2020,Okugawa2020,Fu2021,Kim2021,Palacios2021,Zou2021,Zhang2021,Ghorashi2021,Ghorashi2021a,Shang2022,Li2022a,Zhu2022}.
Some studies proposed to design second-order NHSE from the Benalcazar-Bernevig-Hughes model as a topological quadrupole insulator with corner modes~\cite{Kawabata2020,Okugawa2020,Fu2021,Kim2021}.
This design requires positive couplings along one direction and negative couplings along opposite direction, clearly exotic features not easy to be realized in experiment.
Hybrid skin-topological modes represents a second type of second-order NHSE. It can be realized in experiment but still requires specially designed asymmetric couplings,  limiting its possible extensions~\cite{Lee2019,Li2020,Zou2021,Shang2022}.
Another kind of hybrid skin-topological modes without asymmetric couplings has been proposed recently by adding gain/loss to Chern insulators or anomalous Floquet topological insulators~\cite{Li2022a,Zhu2022}.
Up to now, all the constructions are based on tight-binding lattices.  To further explore second-order NHSE for possible applications, we shall explore in this work possible second-order NHSE in continuous systems, specifically in photonic crystals.

Photonic crystals, whose material parameters and geometric structure can be easily tuned, have already been proven to be a good platform to study different topological states~\cite{Lu2014,Khanikaev2017,Xie2018,Ozawa2019,Smirnova2020,Ota2020,Kim2020,Xue2021}.
Actually the classical wave analogy of Chern insulator was first proposed and realized in gyromagnetic photonic crystals~\cite{Haldane2008,Wang2008,Wang2009}. The Floquet topological insulator and Weyl semi-metal were all first realized in photonic crystals even earlier than their electric counterparts~\cite{Rechtsman2013,Lu2015}.
More importantly, photonic crystals are also a good platform to study non-Hermitian physics where the loss is ubiquitous from material absorption or mode leaking and the gain can be obtained from electrical or optical pumping~\cite{El-Ganainy2018,Miri2019}.
There are also great efforts to study non-Hermitian photonic topological states, mainly in one-dimensional systems~\cite{Malzard2015,Weimann2016,Pan2018,Takata2018,Luo2019,Ao2020,Parto2020,Zhou2020,Song2020,Xia2021,Shi2022}.  However there is less work on non-Hermitian photonic Chern insulators using two-dimensional photonic crystals ~\cite{Silveirinha2019,Li2019,Cerjan2019,Teo2022}.  Our study here shall stimulate further interest in non-Hermitian Chern insulator phases in two-dimensional photonic crystals as continuous systems. 

Studides of topological states and non-Hermitian physics in photonic crystals may be relevant to a wide variety of applications, such as topological delay line, high sensitivity sensors and topological lasers~\cite{Hafezi2011,Hodaei2017,Zeng2019,Harari2018,Bandres2018,Zeng2020}.  Indeed,
topological defects as one way to localize light has attracted much  attention for the possibility to enhance light-matter interactions~\cite{Xie2018a,Noh2018,Chen2019,Zhang2020a}.
NHSE as a new way to localize light has also been studied in photonic crystals most recently~\cite{Zhu2020,Song2020a,Weidemann2020,Zhong2021,Yan2021,Yokomizo2021}.
However the second-order version of NHSE featured in this work as photonic corner skin modes is still not investigated until now. 

In this paper, we propose one way to obtain photonic corner skin modes by adding pseudo inversion symmetric gain/loss to gyromagnetic photonic crystals, which supports chiral edge states.
The spectrum of the edge states is complex under mixed boundary condition and real under OBC. 
Correspondingly, the eigen field is extended along the edge under mixed boundary condition and localized at the corner under OBC.
Similar to the bipolar NHSE from twisted spectral winding numbers in one-dimensional system~\cite{Song2019,Wang2021,Zhang2021a,Li2022}, we observe that part of the chiral edge states are localized at one corner and others are localized at opposite corner.  This is so even though the spectrum of the edge states does not form a loop due to quantum anomaly \cite{Haldane2008}.
Such intriguing phenomena exist in both square lattice and triangle lattice with pseudo inversion symmetry. Furthermore,
by constructing a unit cell preserving pseudo six-fold rotation symmetry, it is found that  the corner skin modes can be simultaneously localized at multiple corners in a triangle lattice.
Our work clearly shows that features of non-Hermitian photonic Chern insulators are sensitive to their boundary conditions.

This work is organized as follows. In Section II, we review the Hermitian gyromagnetic photonic crystal which supports topological chiral edge states in both square lattice and triangle lattice.
In Section III, we first provide a general understanding of the formation of corner skin modes by accumulation of chiral edge states with gain/loss.
Then we discuss the concrete behavior of corner skin modes in various lattices with different designs of gain/loss.
Section IV concludes this work.
All the simulations are performed in COMSOL Multiphysics and we only consider the transverse-magnetic modes.
We use the sign convention of RF modulo in COMSOL, where the positive (negative) imaginary parts represent loss (gain).

\section{Topological chiral edge states in photonic crystals without non-Hermiticity}

\begin{figure}[htbp]
\includegraphics[width=1\linewidth]{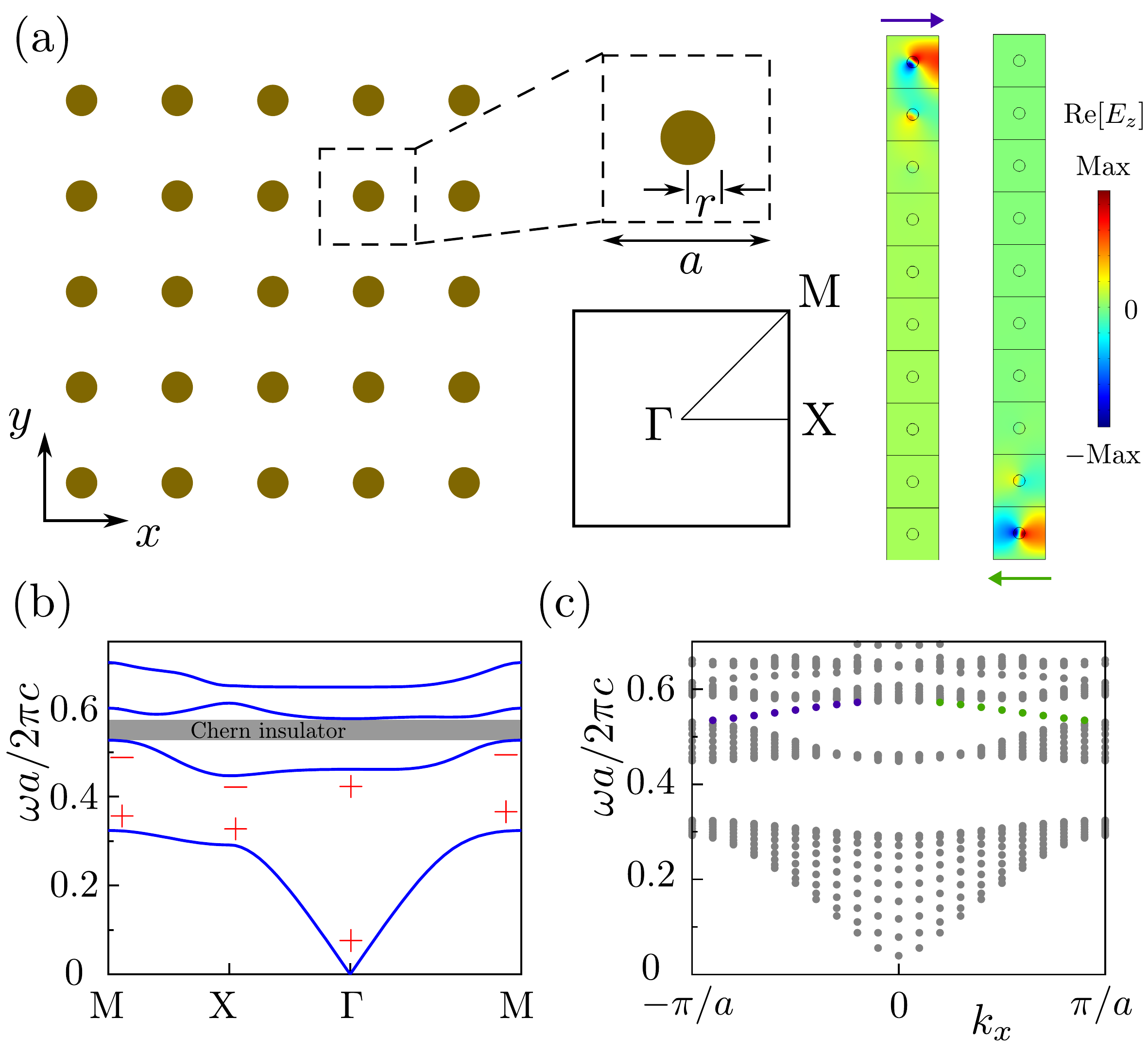}
\caption{
Topological chiral edge state of gyromagnetic photonic crystal in a square lattice.
(a) Two-dimensional photonic crystal in a square lattice is composed of gyromagnetic rods of radius $0.11a$, where $a$ being the lattice constant. The background is air.
The unit cell and first BZ are shown at right.
(b) The bulk bands along high symmetric line in first BZ.
The parity symmetries at high symmetric momentum points are marked for the lower two bands. $+$ ($-$) indicates even (odd) parity of the eigenstates.
The second band gap is a Chern insulator.
(c) The projected band structure as a function of $k_x$. In calculation, periodic boundary condition (PBC) is used along $x$ and OBC is used along $y$. OBC is realized by setting the boundary as perfect electric conductors.  The second band gap supports chiral edge states, one (colored by purple) propagating to right at the upper boundary and the other one (colored by green) propagating to left at the lower boundary. }
\label{squarehermitian}
\end{figure}

We first introduce the gyromagnetic photonic crystals, which has been used to realize reflection-free topological edge states~\cite{Wang2008,Poo2011,Skirlo2014,Liu2020}.
One example is shown in Fig.~\ref{squarehermitian}(a), where the rods made of gyromagnetic materials are regularly distributed in a square lattice and the background is air.
The parameters are the same as Ref.~\cite{Wang2008} with the lattice constant being $a$, the radius of the rods being $r=0.11a$ and the relative permittivity of the rods being $\epsilon_r=15$. The time-reversal symmetry is broken by magnetic filed along the  $z$ direction and the corresponding relative permeability is a tensor
\begin{equation}\label{eq1}
  \mu_r=\left(\begin{array}{ccc}
                14 & 12.4i & 0 \\
                -12.4i& 14 & 0 \\
                0& 0 & 1
              \end{array}
  \right)
\end{equation}
From the unit cell in Fig.~\ref{squarehermitian}(a), one can see that the system preserves 4-fold rotation symmetry which contains the inversion symmetry with $\epsilon_r(\mathbf{r})=\epsilon_r(-\mathbf{r})$ and $\mu_r(\mathbf{r})=\mu_r(-\mathbf{r})$.
The first Brillouin zone is plotted in Fig.~\ref{squarehermitian}(a). $\Gamma$, $\mathrm{X}$ and $\mathrm{M}$ are the high symmetric momentum points.

The bulk band structure along high symmetric line in the first Brillouin zone is studied and shown in Fig.~\ref{squarehermitian}(b).
Usually, the topological properties of the bulk band can be described by the Chern number defined from the Bloch states over the whole Brillouin zone~\cite{Wang2008,Wang2019}.
Here we quickly check the topological properties by considering the inversion symmetry eigenvalues at inversion symmetric momentum points.
From Fig.~\ref{squarehermitian}(b), it is seen that the inversion symmetry eigenvalues are all positive for the first band and possesses an odd number of positive eigenvalues for the second band.  This indicates that the second band gap is a Chern insulator.
Such results are confirmed by the projected band structure shown in Fig.~\ref{squarehermitian}(c), where right propagating chiral edge states (colored by purple) are localized at the upper edge and left propagating chiral edge states (colored by green) are localized at the lower edge.

\begin{figure}[htbp]
\includegraphics[width=\linewidth]{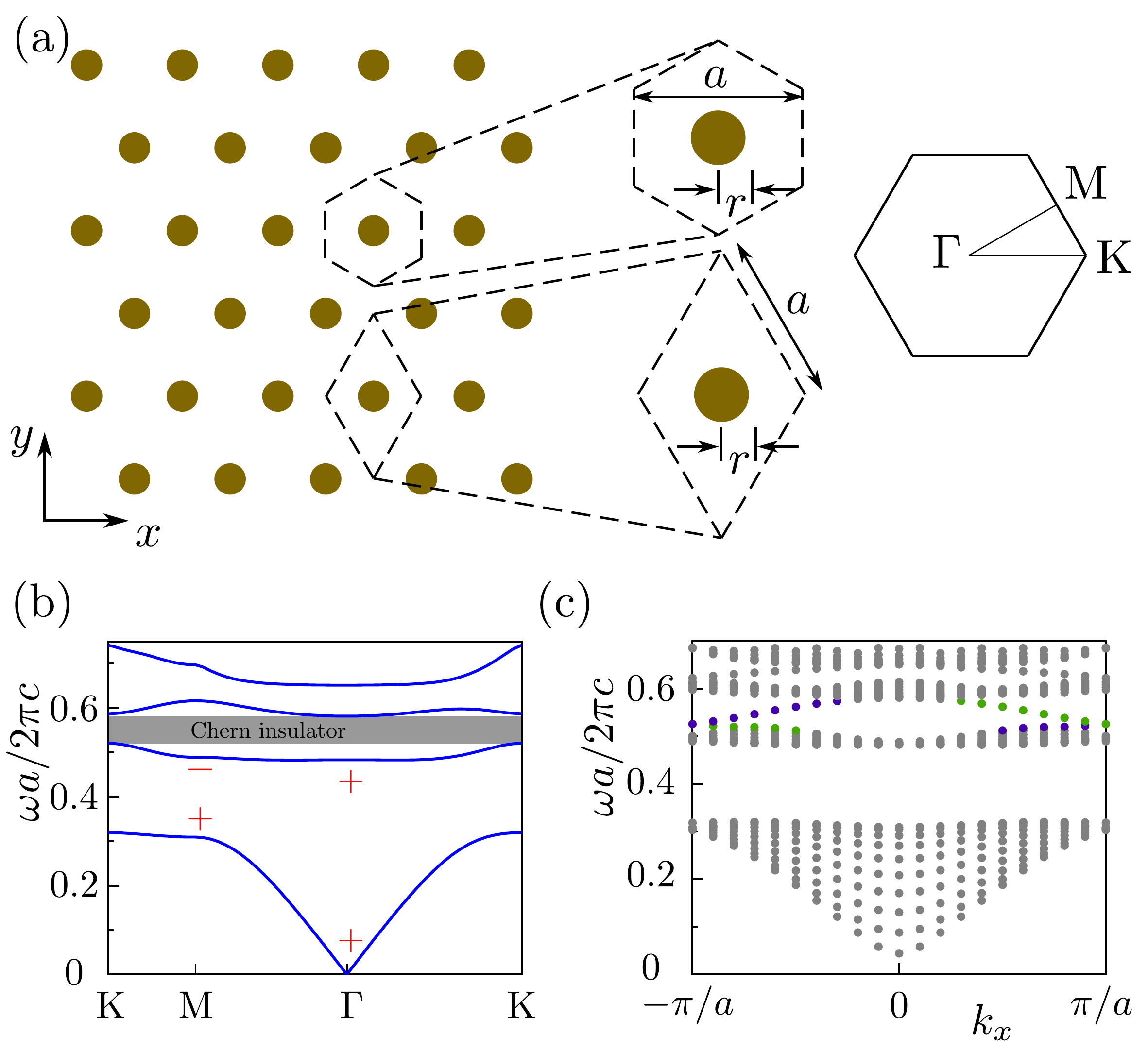}
\caption{Same as in Fig.~\ref{squarehermitian}, except for triangle photonic crystal.
(a) Structure of the photonic crystal in a triangle lattice.
(b) The bulk bands along a high symmetric line.
(c) The projected band structures.
}
\label{trianglehermitian}
\end{figure}

We carry out a parallel study with the  same system parameters but in a triangle lattice as shown in Fig.~\ref{trianglehermitian}(a).
The unit cell can be chosen as hexagon which preserves six fold rotation symmetry or rhombus which preserves inversion symmetry.
Different from the square lattice case, the high symmetric momentum points for the triangle lattice are $\Gamma$, $\mathrm{M}$ and $\mathrm{K}$.
In addition, only $\Gamma$ and $\mathrm{M}$ are inversion symmetry invariant momentum points.
The bulk band structure is shown in Fig.~\ref{trianglehermitian}(b), which is quite similar to the square lattice case due to that the triangle lattice can be obtained by a slightly geometry transformation from the square lattice.
As the square lattice, the inversion symmetry eigenvalues are all positive for the first band and possesses an odd number of positive eigenvalues for the second band,  indicating again that the second band gap is a Chern insulator.
Indeed, the chiral edge states can be observed in the second band gap of projected band structure as shown in Fig.~\ref{trianglehermitian}(c).

Because the right propagating edge states and left propagating edge states are spatially separated (they are orthogonal to each other), we can describe the topological edge states shown in Figs.~\ref{squarehermitian}(c) and ~\ref{trianglehermitian}(c) by a diagonal Hamiltonian under the basis of upper edge state and lower edge state $(\psi_{\mathrm{up}},\psi_{\mathrm{down}})^{T}$. The Hamiltonian of the edge states can be described by

\begin{equation}\label{eq2}
  H_{\mathrm{edge}}=\omega_r\sigma_0+v(k_x-k_r)\sigma_z
\end{equation}
Here $\omega_r$ being a reference angular frequency in the middle of the second band gap. $k_r$ is a reference momentum. $v$ is the group velocity.
$v>0$ in our system, meaning the upper (lower) edge state propagate to right (left).
$\sigma_0$ is two by two identity matrix and $\sigma_z$ is the third Pauli matrix.
This Hamiltonian of the edge states is useful for the understanding of photonic corner skin modes later.

\section{Photonic corner skin modes in non-Hermitian photonic crystals with pseudo-inversion symmetry}

\subsection{General understanding of photonic corner skin modes}

\begin{figure}[htbp]
\includegraphics[width=0.8\linewidth]{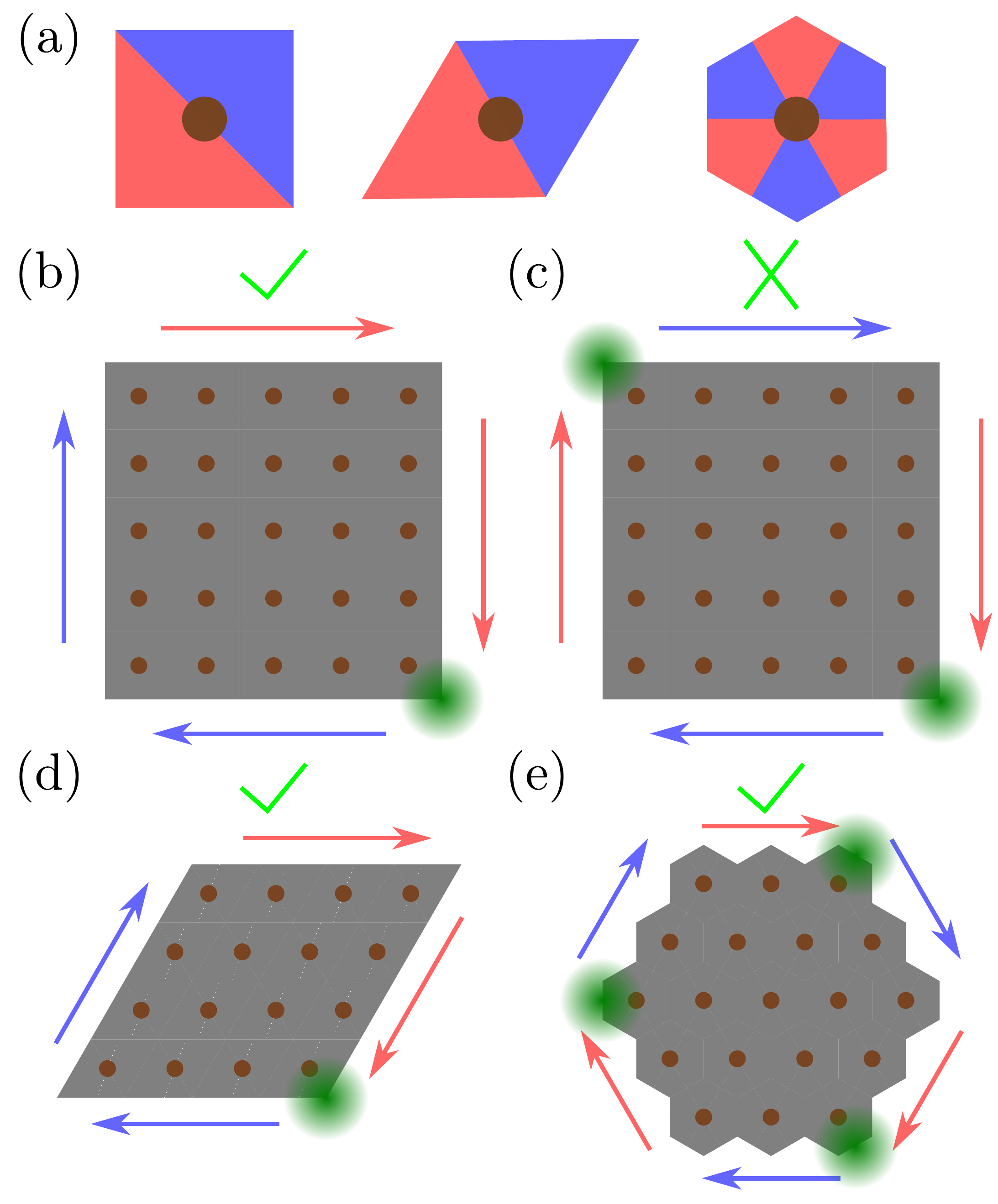}
\caption{Schematic for corner skin modes in non-Hermitian photonic crystals with pseudo inversion symmetry.
(a) Three examples of the unit cell with pseudo inversion symmetry. The red (blue) part indicates gain (loss).
(b) One possible configuration for the chiral edge states localized at one corner.
(c) One pseudo inversion symmetry forbidden configuration where the chiral edge states are localized at two corners.
(d)(e) Two possible configurations for the chiral edge states localized at one or three corners.}
\label{schematic}
\end{figure}

The photonic Chern insulators are quite robust even when the system has non-Hermiticity.
With non-Hermiticity, the topological chiral edge states are still there without the band gap closing~\cite{Silveirinha2019,Li2019,Cerjan2019}.
We introduce gain and loss to the unit cell as some examples in Fig.~\ref{schematic}(a), where the red area represents gain and blue area represent loss.
The unit cell still preserves pseudo-inversion symmetry with $\epsilon_r(\mathbf{r})=\epsilon_r^\dagger(-\mathbf{r})$ and $\mu_r(\mathbf{r})=\mu_r^\dagger(-\mathbf{r})$.
As a general analysis, note that if the chiral edge states contain gain (loss) in the upper edge, then we can conclude that the chiral edge states in the lower edge contains loss (gain) due to the pseudo-inversion symmetry.
The Hamiltonian of the edge states is then modified as
\begin{equation}\label{eq3}
  H^{m}_{\mathrm{edge}}=\omega_r\sigma_0+v(k_x-k_r)\sigma_z+i\gamma\sigma_z
\end{equation}

The gain/loss can be treated as a modification to the quasi momentum $\tilde{k}_x=k_x+i\gamma/v$, which make the waves localize to right (left) for $\gamma<0$ ($\gamma>0$) due to NHSE under OBC.
Different from the usual NHSE in 1D chain, here the spectrum of the edge state does not form a loop so that it cannot be described by spectral winding numbers but by the sign of $\gamma$.

Figs.~\ref{schematic}(b)-\ref{schematic}(e) show some configurations of the chiral edge states with gain/loss. The red (blue) arrow represents gain (loss) chiral edge states.
Gain (loss) chiral edge states accumulate waves to (opposite to) the propagation direction.
Fig.~\ref{schematic}(b) is one possible configuration for square lattice where the waves are localized at the right-down corner due the accumulation effect.
Fig.~\ref{schematic}(c) shows the configuration where the waves are localized at two opposite corners.
However, the chiral edge states on opposite edge having same gain/loss are forbidden by the pseudo-inversion symmetry.
So for the square lattice case, all the waves are localized at one corner.
For the triangle lattice case however, it is possible to localize wave at one corner or three corners as shown in Figs.~\ref{schematic}(d) and \ref{schematic}(e).

\subsection{Photonic corner skin modes due to bipolar second-order NHSE}

\begin{figure}[htbp]
\includegraphics[width=1\linewidth]{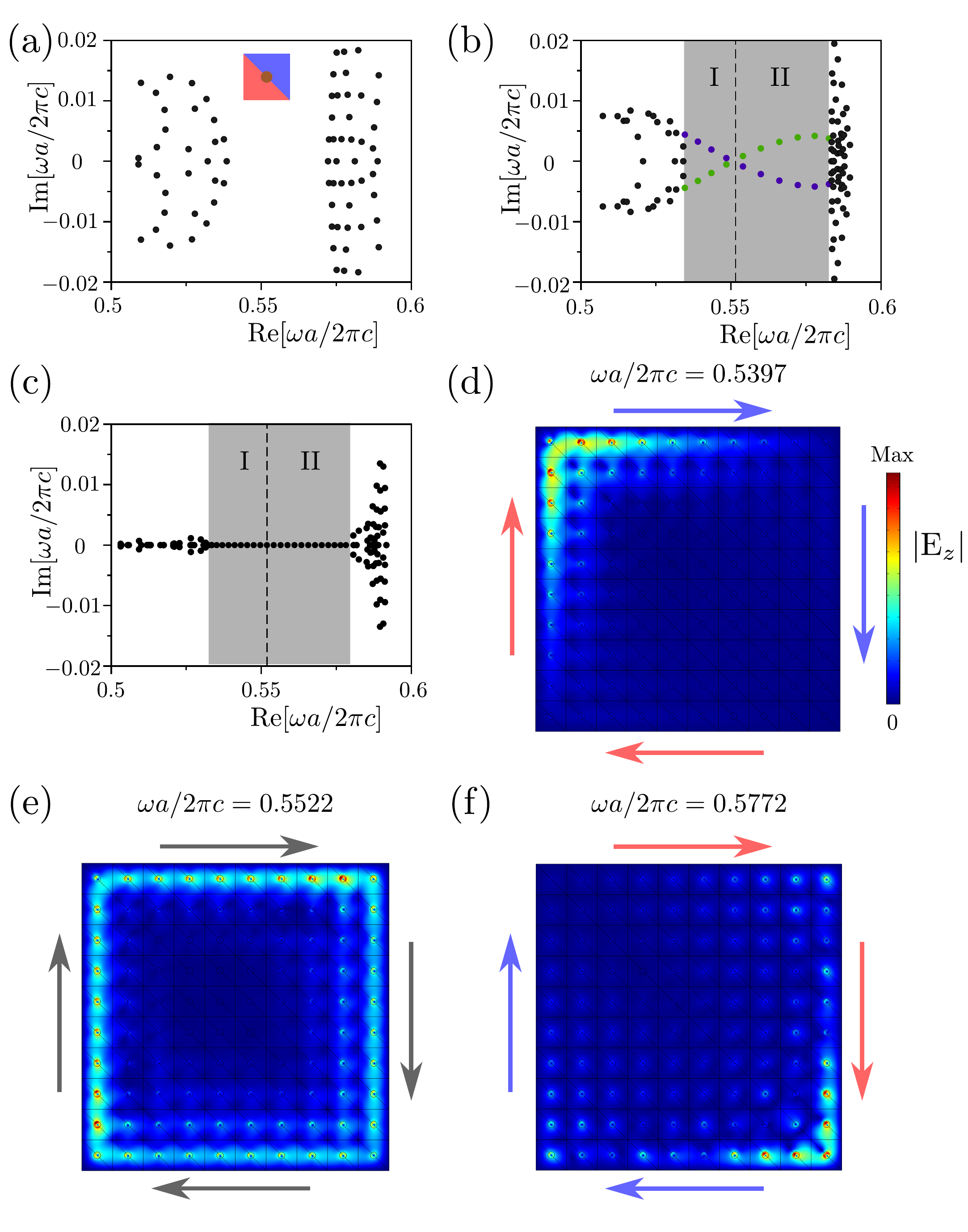}
\caption{Corner skin modes in square lattice with pseudo inversion symmetry.
(a) Spectrum for the system with PBC in both directions. Unit cell is shown in the inset. (b) Spectrum of the system with mixed boundary condition, PBC in one direction and OBC in another direction. The edge states for upper (lower)  boundary are colored by purple (green). (c) Spectrum for the system with OBC in both directions.
(d)(e)(f) The field profiles for the topological edge states in open boundary condition. (d) corner skin modes with lower energy are localized at left-up corner, (e) extended topological edge states at the transition point and (f) corner skin modes with higher energy is localized at right-down corner.}
\label{squareinversion}
\end{figure}

We now study non-Hermitian photonic crystals in a square lattice with pseudo inversion symmetry, whose unit cell is shown in the insert figure of Fig.~\ref{squareinversion}(a).
The gain (loss) is added to the air by tuning the relative permittivity to $\epsilon_r=1-0.8i$ ($\epsilon_r=1+0.8i$).
The spectrum of system with PBC in both directions is studied and shown in Fig.~\ref{squareinversion}(a).
There are two clusters corresponding to the second bulk band and third bulk band.
The line band gap is between two bulk bands.
We then study the spectrum of mixed boundary condition and the results are shown in Fig.~\ref{squareinversion}(b).
It is observed that the bulk band clusters become tighter than that in PBC.
The reason is that the nonreciprocal bulk band combined with the gain/loss makes the system support first-order NHSE, so that the bulk spectrum is sensitive to the boundary condition~\cite{Zhong2021}.
We also see that there are complex chiral edge states across the line band gap.
The spectrum for the system with OBC in both directions is shown in Fig.~\ref{squareinversion}(c).
Again the bulk spectrum becomes tighter.
More importantly, the spectrum of chiral edge states has a dramatic change from complex in the mixed boundary to real.
Such dramatic change is one feature of NHSE.
Specifically, we witness NHSE occuring to topological edge states and can hence be understood as second-order NHSE.

\begin{figure}[htbp]
\includegraphics[width=1\linewidth]{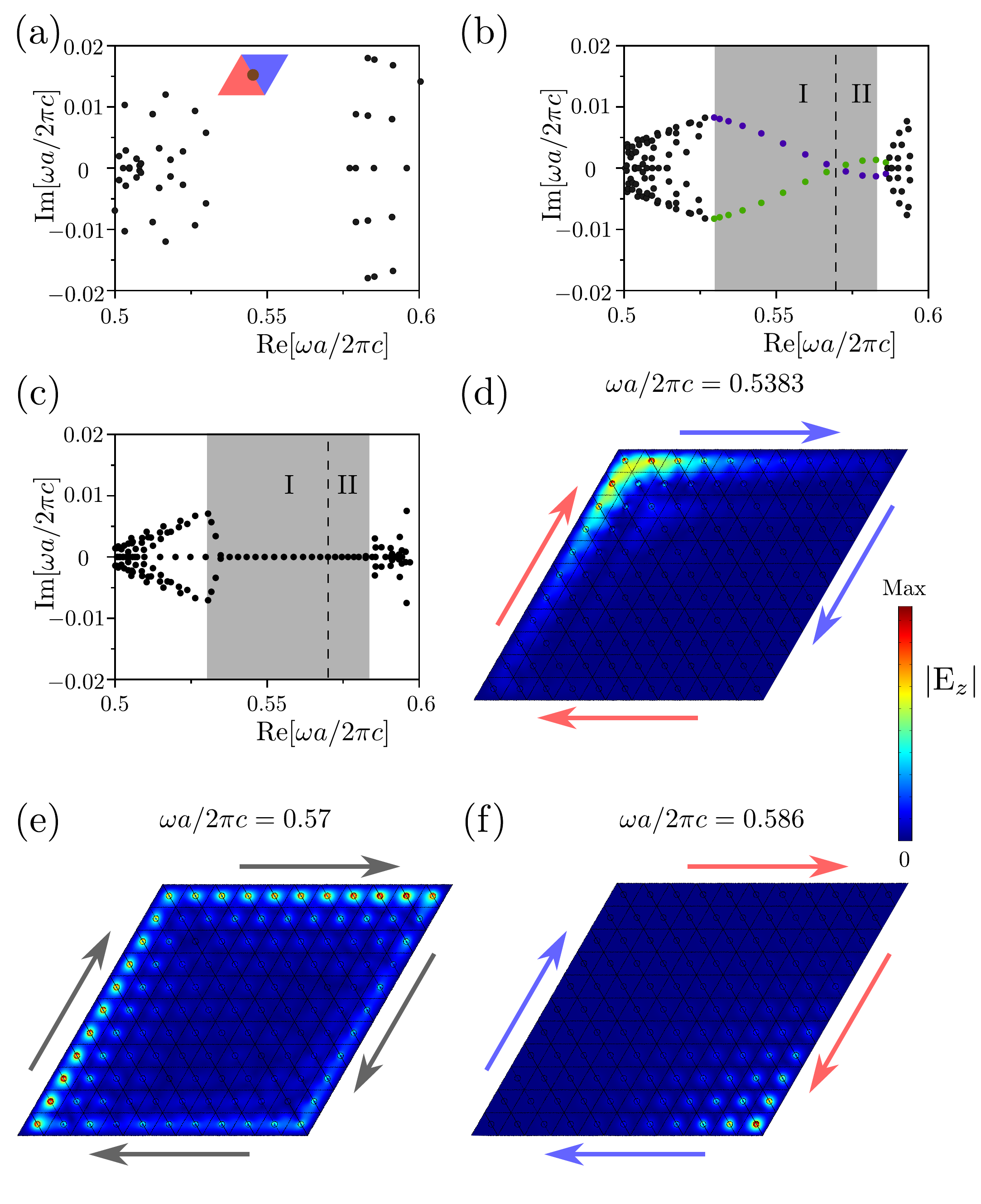}
\caption{Same as in Fig.~\ref{squareinversion}, except for triangle photonic crystal.
(a) The spectrum for the system with PBC in both directions. (b) The spectrum for the system with mixed boundary condition. (c) The spectrum for the system with OBC in both directions. (d)(e)(f) The field profiles are shown to illustrate the transition between different localization behavior of corner modes.}
\label{triangleinversion}
\end{figure}

We focus on the spectrum of edge states in the second band gap.
In the mixed boundary condition, the spectrum of edge states can be split into two parts as shown in Fig.~\ref{squareinversion}(b): in first part marked by I, the up (down) edge states has loss (gain) corresponding to $\gamma>0$ and in second part marked by II, the up (down) edge states has gain (loss) corresponding to $\gamma<0$.
According to Eq.~\ref{eq3}, these two parts have different localization behavior in OBC.
The eigenfields with OBC are shown in Figs.~\ref{squareinversion}(d)-\ref{squareinversion}(f).
It is seen that for part I the waves are localized at left-up corner shown in Fig.~\ref{squareinversion}(d), at the transition point the waves are extended along the edge shown in Fig.~\ref{squareinversion}(e) and for  part II the waves are localized at right-down corner shown in Fig.~\ref{squareinversion}(f).
Such phenomena called bipolar NHSE has been observed in 1D system and are usually connected with a twisted spectral winding~\cite{Song2019,Wang2021,Zhang2021a,Li2022}.
Here we observe the similar phenomena in a 2D system, which can be understood as a bipolar second-order NHSE.

Next we examine a similar photonic crystal but in a triangle lattice configuration. The unit cell is shown in the insert figure of Fig.~\ref{triangleinversion}(a).
Similar phenomena have been observed. Compared with the square lattice case, the bulk spectrum is less sensitive to the boundary conditions as shown in Figs.~\ref{triangleinversion}(a)-\ref{triangleinversion}(c).
The spectrum of the edge states is however  still sensitive to the boundary condition, which are complex in mixed boundary condition and real in OBC as shown in Figs.~\ref{triangleinversion}(b) and \ref{triangleinversion}(c).
There are also two parts, part I corresponding to $\gamma>0$ and part II corresponding to $\gamma<0$.
In part I, the waves are localized at left-up corner as shown in Fig.~\ref{triangleinversion}(d), at phase transition point the waves are extended along the edge as shown in Fig.~\ref{triangleinversion}(e) and in part II the waves are localized at right-down corner as shown in Fig.~\ref{triangleinversion}(f).

\subsection{Multiple-corner skin modes}

\begin{figure}[htbp]
\includegraphics[width=1\linewidth]{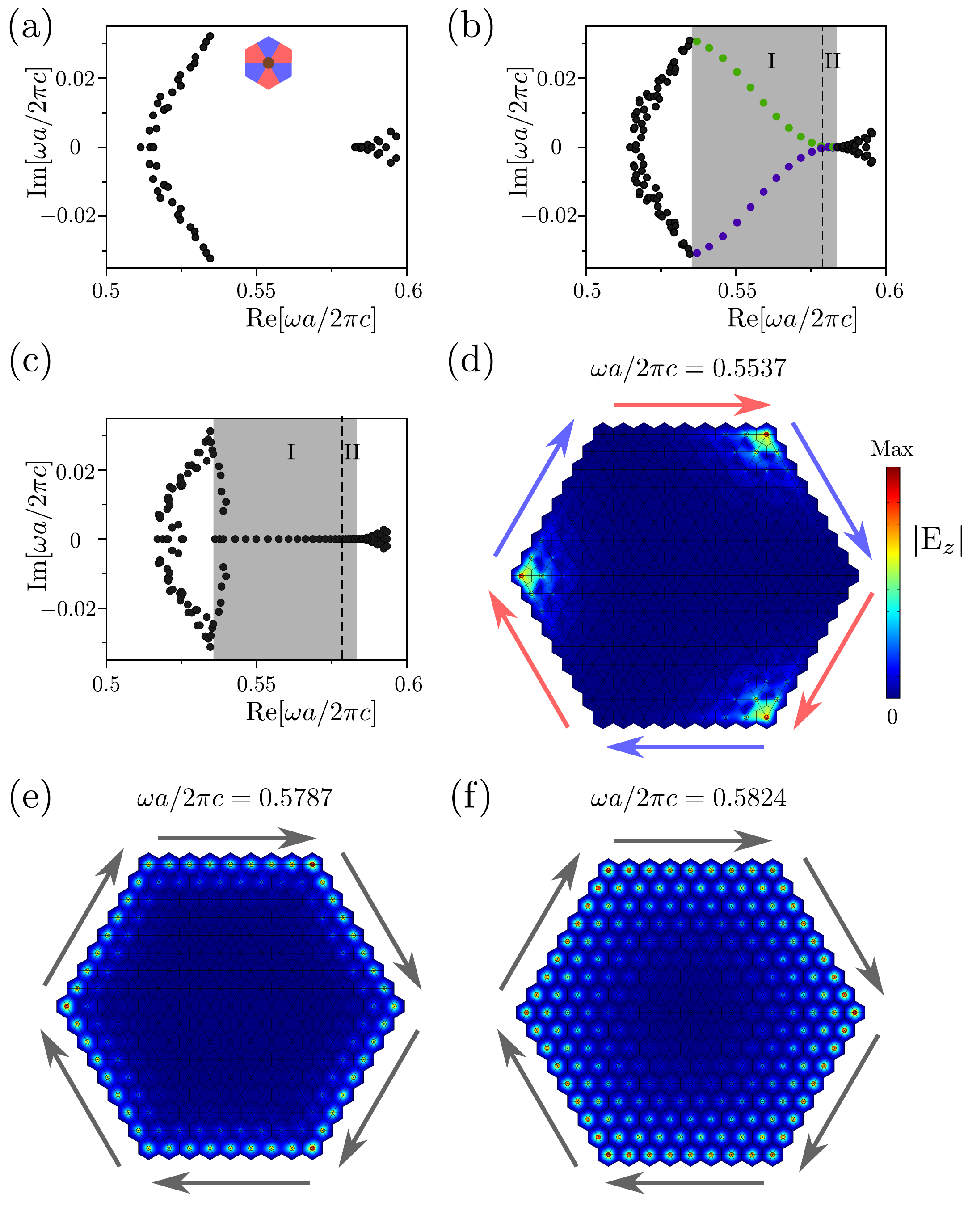}
\caption{Corner skin modes in a triangle lattice configuration with pseudo $C_6$ symmetry.
(a) The spectrum of the system with PBC in both directions. (b) The spectrum of the system with mixed boundary condition. (c) The spectrum for the system with OBC in both directions.
(d) The field profiles for one corner skin modes where the fields are localized at three corners.
(e)(f) Two examples of extended topological edge states.
}
\label{trianlgeC6}
\end{figure}

Previous studies show the photonic corner skin modes are localized at one corner.
Next we provide a case where the photonic corner skin modes are localized at multiple corners.
The unit cell is shown in insert figure of Fig.~\ref{trianlgeC6}(a).
The unit cell preserves pseudo six fold rotation symmetry with $\epsilon_r(\mathbf{r})=\epsilon^\dagger_r(C_6\mathbf{r})$ and $\mu_r(\mathbf{r})=\mu^\dagger_r(C_6\mathbf{r})$, which includes the pseudo inversion symmetry with $\epsilon_r(\mathbf{r})=\epsilon^\dagger_r(-\mathbf{r})$ and $\mu_r(\mathbf{r})=\mu^\dagger_r(-\mathbf{r})$.
The spectrum for the system with PBC in both directions is shown in Fig.~\ref{trianlgeC6}(a).
As in the previous case, there are two bulk band clusters.
The spectrum for the system with mixed boundary condition and OBC in both directions are show in Figs~\ref{trianlgeC6}(b) and \ref{trianlgeC6}(c).
Obviously, the bulk spectrum is almost unchanged in different boundary conditions.  The edge spectrum is again quite sensitive to the boundary condition insofar as the edge spectrum is complex under mixed boundary condition and real under OBC.
Focusing on the spectrum of the edge states, we again see that they still can be split into two parts, part I and part II.
Under mixed boundary condition, the spectrum of edge states are complex in part I corresponding to $\gamma>0$ and real in part II corresponding to $\gamma=0$.
According to previous analysis, the waves are localized in part I and extended in part II when we choose OBC in both directions.
Such results are confirmed by the field profiles.
For states belonging to part I, the waves are localized at three corners as shown in Fig.~\ref{trianlgeC6}(d).
For part II, the waves are extended as shown in Figs.~\ref{trianlgeC6}(e) and \ref{trianlgeC6}(f).

\section{Conclusion and discussion}

In this article we have presented an innovative way to localize light by adding designed gain/loss to photonic crystals with chiral edge states.
The light is localized at corners due to second-order NHSE, subject to particular symmeries in the gain/loss introduced to the lattice. 
Specifically, photonic corner skin modes due to bipolar second-order NHSE and multiple-corner skin modes are predicted in continuous systems.   
The resulting interesting localization behavior of light should be of experimental interest and may be used to enhance light-matter interactions.
Compared with the better known first-order NHSE, our results clearly demonstrate that second-order NHSE can be engineered by use of crystal symmetries. Photonic crystals are hence identified as a verstile platform to investigate and exploit second-order NHSE.



%

\end{document}